\documentclass[a4paper]{jpconf}%
\usepackage{amssymb}
\usepackage{graphicx}
\usepackage{amsmath}
\usepackage{amsfonts}%
\begin{document}

\title{A study of the resonances $K_{0}^{*}(800)$ and $K_{0}^{*}(1430)$}
\author{Milena So{\l }tysiak$^{1,*}$, Thomas Wolkanowski$^{2}$ and Francesco Giacosa$^{1,2}$ \address{$^1$Institute of Physics, Jan Kochanowski University, PL-25406 Kielce, Poland $^2$Institute for Theoretical Physics, Goethe University, D-60438 Frankfurt am Main, Germany} }

\begin{abstract}
We study the scalar kaonic states $K_{0}^{\ast}(800)$ and $K_{0}^{\ast}(1430)$
by using a relativistic QFT Lagrangian in which only a single kaonic field
corresponding to the well-established scalar state $K_{0}^{\ast}(1430)$ is
considered and in which both derivative and non-derivative interaction terms
are taken into account. Even if the scalar spectral function shows a unique
peak close to $1.4$ GeV, we find two poles in the complex plane:
$1.413\pm0.002-i(0.127\pm0.003)$ GeV, which is related to the seed
quark-antiquark state $K_{0}^{\ast}(1430),$ and $0.746\pm0.019-i(262\pm0.014)$
GeV, which is an additional companion pole related to $K_{0}^{\ast}(800)$. As
a further investigation for increasing $N_{c}$ confirms, $K_{0}^{\ast}(800)$
emerges as a dynamically generated four-quark object as a consequence of
pion-kaon loops.

\end{abstract}

\ead{\\$^*$ milena.soltysiak@op.pl}

\section{Introduction}

The scalar resonance $K_{0}^{\ast}(1430)$ with $I(J^{P})=\frac{1}{2}(0^{+})$
is a well-established state listed in the summary table of the Particle Data
Group (PDG) \cite{pdg}. On the contrary, the scalar state $K_{0}^{\ast}(800)$,
also with $I(J^{P})=\frac{1}{2}(0^{+})$ and known as $\kappa$ , still needs
confirmation. The acceptance of the lightest $\kappa$ in the summary table of
the PDG is important since it would allow to complete the nonet of the light
scalar states below $1$ GeV.\newline The main aim of this work is to
understand the nature of the two resonances $K_{0}^{\ast}(1430)$ and
$K_{0}^{\ast}(800)$ using the quantum field theoretical model described in
Ref. \cite{our} (on which this proceedings is based). In particular, we shall concentrate on the existence and the properties of the latter state. The key ingredient of our theoretical approach is to consider at the same time both derivative and non-derivative interaction terms, see Sec. 2. We perform a fit of the four parameters of our model to the measured data points of $\pi K$ phase shift in the $I=1/2,J=0$ channel from Ref. \cite{Aston:1987ir} (See Sec. 3.1) and
obtain a remarkably good agreement with the experiment. As a consequence of
the fit, we determine the scalar kaonic spectral function up to $1.8$ GeV, in
which only one single peak corresponding to $K_{0}^{\ast}(1430)$ appears.
Namely, there is no peak corresponding to $K_{0}^{\ast}(800)$ but only a small
enhancement in the corresponding energy region. Next, we calculate the
coordinates of the poles on the complex plane. Despite the fact that we
included only one \textquotedblleft seed\textquotedblright\ state in the
effective Lagrangian, we find two poles: one corresponding to the peak in the
spectral function at $1.4$ GeV and thus to $K_{0}^{\ast}(1430)$, and one
corresponding to the light $\kappa$. We get additional information about the
nature of both resonances by studying the changes of the spectral function and
the movement of the poles for increasing $N_{c}$: $K_{0}^{\ast}(1430)$ behaves
as a $q\bar{q}$ state, while the light $\kappa$ is a dynamically generated
state and as such a non-ordinary meson (see e.g.
Refs. \cite{dullemond,black,oller,oller2,pelaez,fariborz} and refs. therein).

To confirm the correctness of our model, we study in Sec. 3.2 variations of it
in which only the non-derivative or only the derivative interaction term is
taken into account, respectively. We also studied variations of the form
factor. In all cases we obtain a worse agreement with the experiment with
respect to Sec. 3.1. It turns out that using both derivative and
non-derivative terms is crucial to obtain a good description of experimental
data. Finally, we summarize our results in Sec. 3.4.

\section{The model}

We use a relativistic interaction Lagrangian consisting of both derivative and
nonderivative terms for a single $q\bar{q}$ scalar state denoted as
$K_{0}^{\ast}$:
\begin{equation}
\mathcal{L}_{int}=aK_{0}^{\ast+}K^{-}\pi^{0}+bK_{0}^{\ast+}\partial_{\mu}%
K^{-}\partial^{\mu}\pi^{0}+\ldots\text{ ,} \label{lag}%
\end{equation}
where dots refer to complex conjugation and other members of the isospin
multiplet. The model is naturally obtained as a piece of more complete chiral
models, e.g. Ref. \cite{elsm}. The decay width is given by:
\begin{equation}
\Gamma_{K_{0}^{\ast}}(m)=3\frac{\left\vert \vec{k}_{1}\right\vert }{8\pi
m^{2}}\left[  a-b\frac{m^{2}-M_{K}^{2}-M_{\pi}^{2}}{2}\right]  ^{2}F_{\Lambda
}(m)\text{ ,}%
\end{equation}
where the pion and kaon masses are denoted as $M_{\pi}$ and $M_{K}$, and
$\vec{k}_{1}$ is the three-momentum of the pion. The form-factor $F_{\Lambda
}(m)=e^{-2\left\vert \vec{k}_{1}\right\vert ^{2}/\Lambda^{2}}$, where
$\Lambda$ stays for an energy scale, assures that calculations are finite. The
propagator of $K_{0}^{\ast}$ reads
\begin{equation}
\Delta_{K_{0}^{\ast}}(p^{2}=m^{2})=\left[  m^{2}-M_{0}^{2}+\Pi(m^{2}%
)+i\varepsilon\right]  ^{-1}\text{ ,}%
\end{equation}
where $M_{0}$ is the bare mass of $K_{0}^{\ast}(1430)$ and $\Pi(m^{2}%
)=\mathrm{Re}(m^{2})+i\mathrm{Im}(m^{2})$ is the one-loop contribution with
one kaon and one pion circulating in it. We recall that the optical theorem
implies that $\mathrm{Im}\Pi(m)=m\Gamma_{K_{0}^{\ast}}(m).$ The spectral
function is given by:
\begin{equation}
d_{K_{0}^{\ast}}(m)=\frac{2m}{\pi}|\mathit{Im}\Delta_{K_{0}^{\ast}}%
(p^{2}=m^{2})|\text{ .}%
\end{equation}
The quantity $d_{K_{0}^{\ast}}(m)dm$ is the probability that the mass of the
resonance is between $m$ and $m+dm.$ The normalization condition $\int
_{0}^{\infty}d_{K_{0}^{\ast}}(m)\mathrm{dm}=1$ must hold. For the details of
the mathematical treatment used here, we refer to \cite{thomas}. In the next
section we investigate the form of the spectral function, the poles of the
propagator, and their dependence on the number of colors $N_{c}.$

\section{Results}

\subsection{Fit to the phase shift data}

We consider the $\pi K$ phase shift in the $I=1/2,J^{P}=0^{-}$ channel below
$1.8$ GeV. Under the assumption that the $s$-channel contribution of the field
$K_{0}^{\ast}$ dominates, the phase shift reads
\begin{equation}
\delta(m)=\frac{1}{2}\arccos\left[  1-\pi\Gamma_{K_{0}^{\ast}}(m)d_{K_{0}%
^{\ast}}(m)\right]
\end{equation}
Using the data of Ref. \cite{Aston:1987ir} (see also Ref. \cite{ishida}) we
perform a fit in order to determine four parameters of our model $a,b,$
$M_{0},$ and $\Lambda$, see Tab. 1. The result of the fit is presented in
Fig. 1. We obtain a good agreement with the experimental data points ($\chi
^{2}/d.o.f.=1.25$). \newline

\begin{table}[ptb]
\caption{The value of the parameters of our model.}%
\centering%
\begin{tabular}
[c]{@{}lllllllllllllll}%
\br Parameter & Value &  &  &  &  &  &  &  &  &  &  &  &  & \\
\mr$a$ & $1.60\pm0.22$ GeV &  &  &  &  &  &  &  &  &  &  &  &  & \\
$b$ & $-11.16\pm0.82$ GeV$^{-1}$ &  &  &  &  &  &  &  &  &  &  &  &  & \\
$m_{0}$ & $1.204\pm0.008$ GeV &  &  &  &  &  &  &  &  &  &  &  &  & \\
$\Lambda$ & $0.496\pm0.008$ GeV &  &  &  &  &  &  &  &  &  &  &  &  & \\
\br &  &  &  &  &  &  &  &  &  &  &  &  &  &
\end{tabular}
\end{table}

\begin{figure}[h]
\begin{center}
\includegraphics[width=0.4 \textwidth] {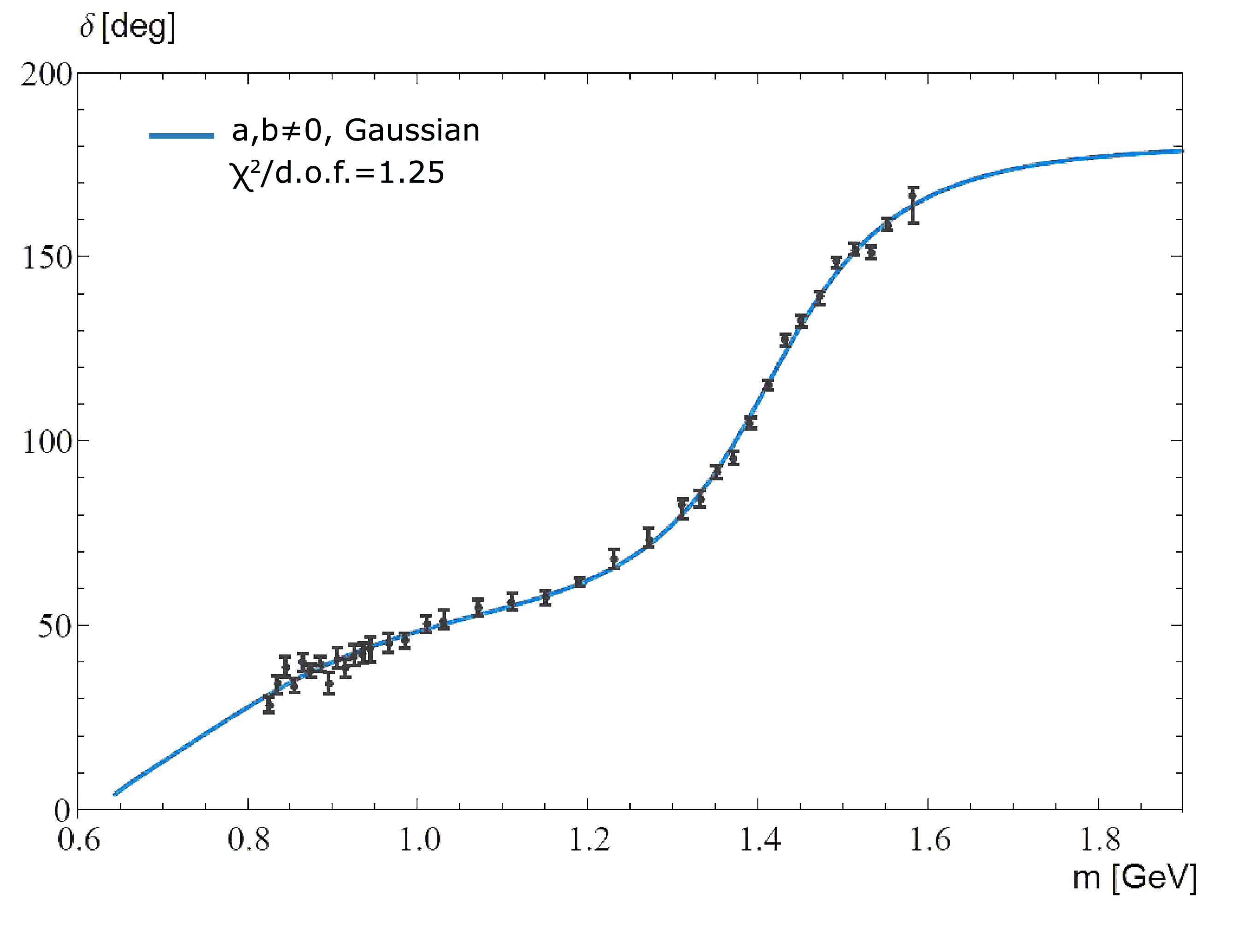}
\end{center}
\caption{Result of the fit to the measured data points of the $\pi K$ phase
shift in $I=1/2,J=0$ channel.}%
\end{figure}According to Eq. (4) and using the parameters reported in Tab. 1,
we report in Fig. 2 the scalar kaonic spectral function up to $1.8$ GeV. We
observe a single peak close to $1.4$ GeV corresponding to $K_{0}^{\ast
}(1430),$ but there is no peak for $K_{0}^{\ast}(800)$ (only a small
enhancement in the low energy is visible). For a comparison with the spectral
function of the narrow Breit-Wigner type vector state $K^{\ast}(892)$ we refer
to \cite{procwarsaw}.

We study the large-$N_{c}$ limit of the spectral function according to the
rescaling $a\longrightarrow\sqrt{3/N_{c}}a$, $b\longrightarrow\sqrt{3/N_{c}}%
b$, where $N_{c}$ is the number of colors, see Fig. 3. (Namely, the Lagrangian
(\ref{lag}) contains a three-leg interaction term, whose amplitude scale as
$1/\sqrt{N_{c}},$ e.g. Ref. \cite{witten}). The peak corresponding to
$K_{0}^{\ast}(1430)$ becomes narrower and higher with decreasing
$\lambda=3/N_{c}$, but the enhancement corresponding to $K_{0}^{\ast}(800)$
becomes smaller and finally disappears for increasing $N_{c}$.

\begin{figure}[h]
\begin{center}
\begin{minipage}{14pc}
\includegraphics[width=14pc]{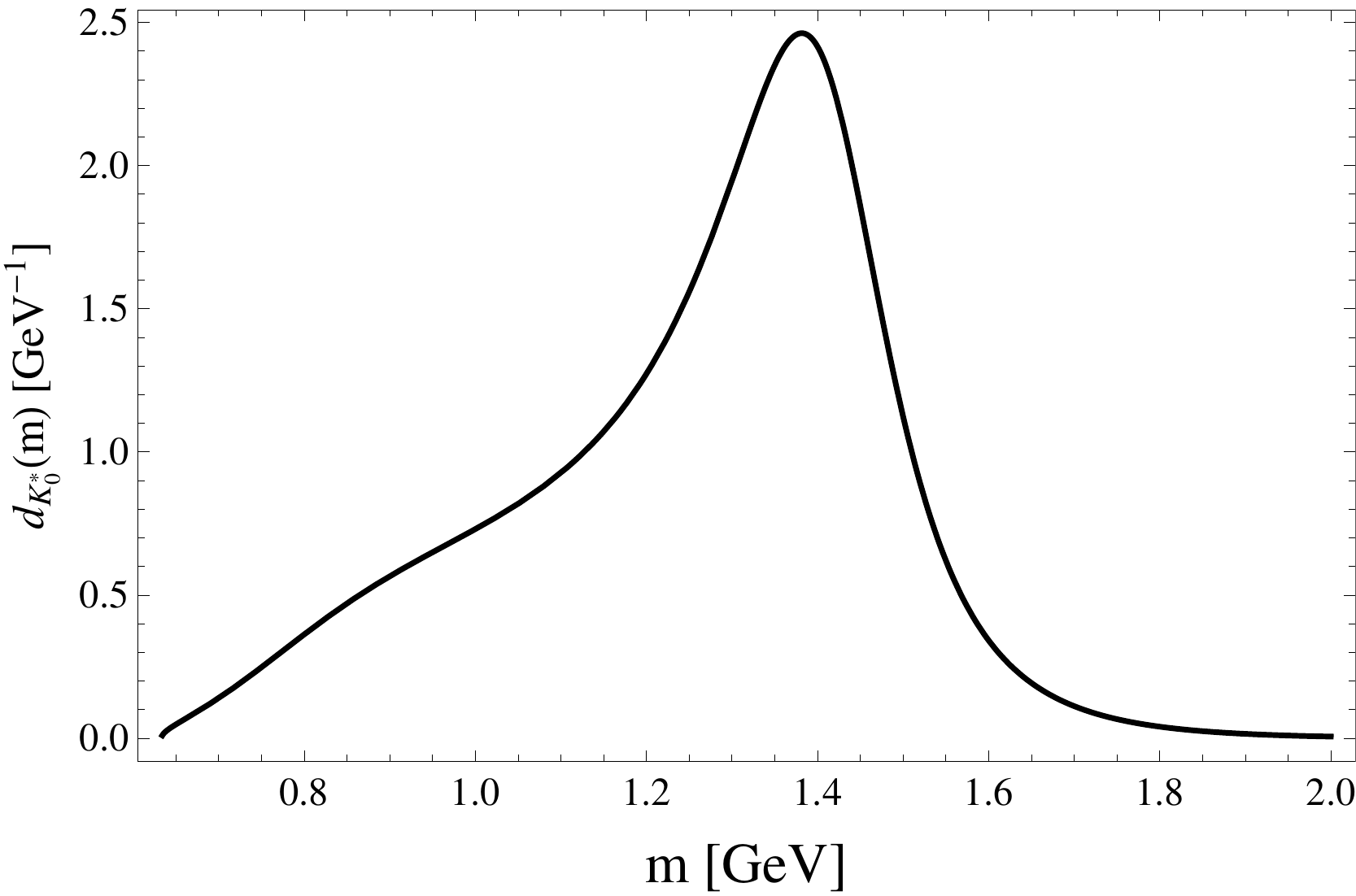}
\caption{Scalar kaonic spectral function from Eq. (4).}
\end{minipage}\hspace{2pc}\begin{minipage}{14pc}
\includegraphics[width=14pc]{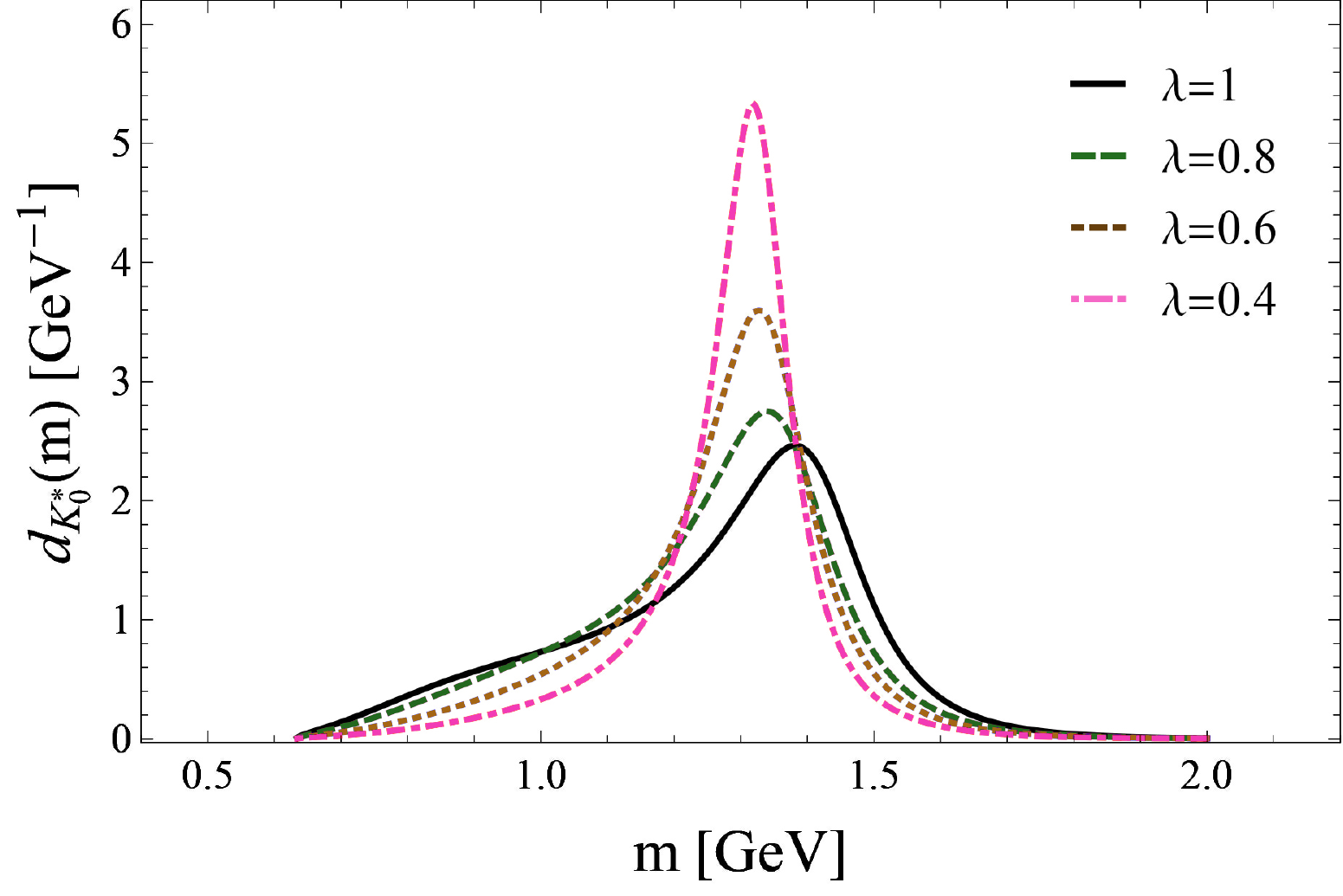}
\caption{Scalar kaonic spectral function studied in large-$N_c$ limit.}
\end{minipage}
\end{center}
\end{figure}

As a next step we determine the positions of the poles in the complex plane.
For the heavier state $K_{0}^{\ast}(1430)$ we get $1.413\pm0.002-i(0.127\pm
0.003)$ GeV. In addition, we also find the lighter state $\kappa$:
$0.746\pm0.019-i(262\pm0.014)$ GeV. In Fig. 5 we show the pole trajectories as
function of $N_{c}$. The pole of $K_{0}^{\ast}(1430)$ moves toward to the real
energy axis, while the pole of $K_{0}^{\ast}(800)$ moves away from it and
disappears for $N_{c}\gtrsim13$.

\begin{figure}[h]
\begin{center}
\includegraphics[width=0.4 \textwidth] {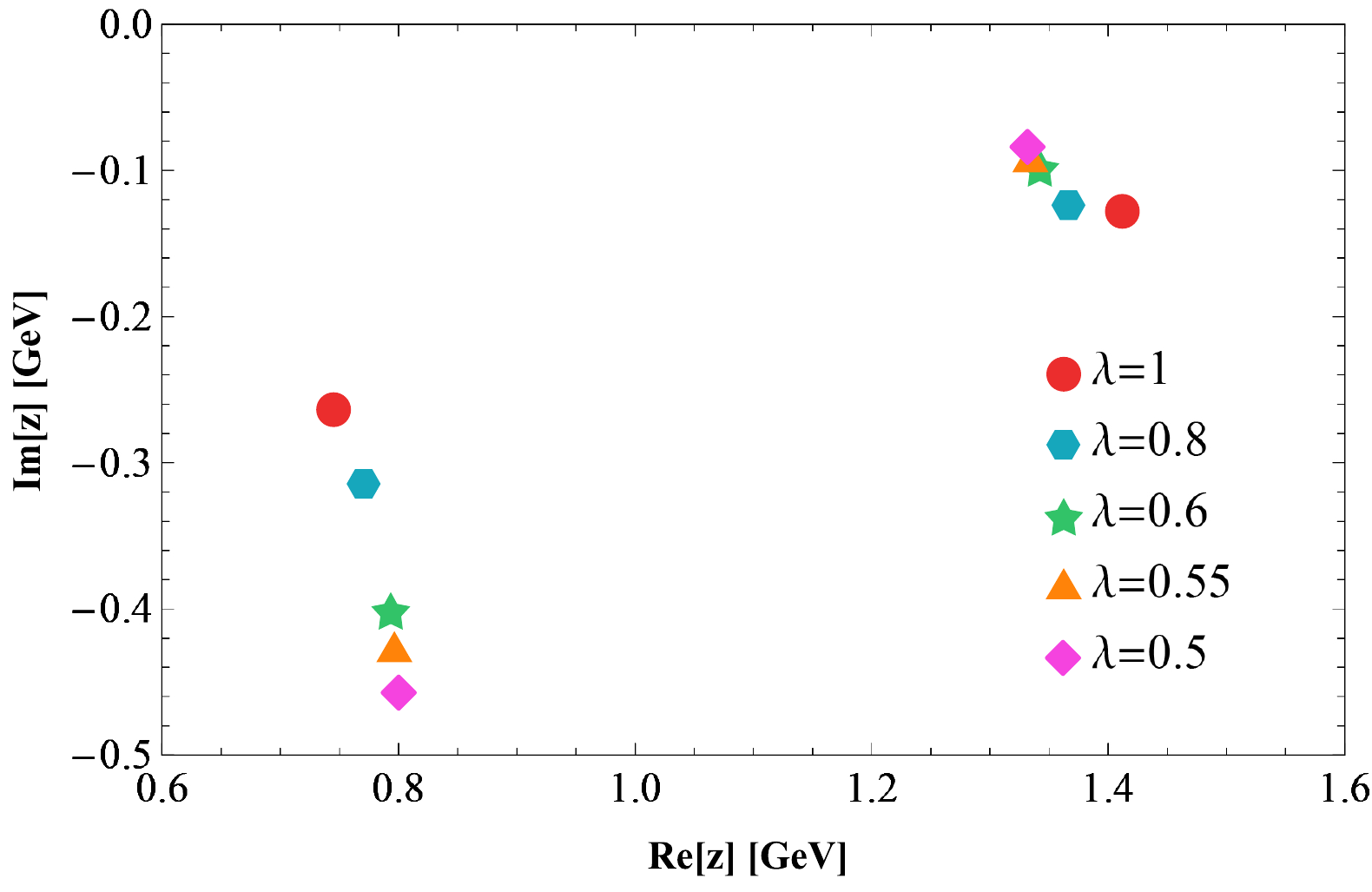}
\end{center}
\caption{Positions of the poles of two resonances $K_{0}^{\ast}(1430)$ and
$K_{0}^{\ast}(800)$ and their trajectories in large-$N_{c}$ limit.}%
\end{figure}

In conclusion, even if in the effective Lagrangian of Eq. (\ref{lag}) a single
$K_{0}^{\ast}$ seed state is considered, we obtain both scalar states
$K_{0}^{\ast}(1430)$ and $\kappa\equiv K_{0}^{\ast}(800)$. Moreover, the
behavior of the spectral function and the pole trajectories as function of
$N_{c}$ show that $K_{0}^{\ast}(1430)$ is a predominantly quark-antiquark
state, while the light $\kappa$ is generated dynamically, by the pion-kaon
loop dressing a preformed quark-antiquark state. Accordingly, it disappears
when the interaction strength is larger than a certain threshold. \newline

\subsection{Modifications of the model}

We study different variations of our model in order to test the robustness of
the results. First, we consider two different limits: one only with the
derivative term ($a\neq0$ and $b=0$ in\ Eq. (\ref{lag})) and one only with the
non-derivative term ($a=0$ and $b\neq0$ in\ Eq. (\ref{lag})). We perform the
fits to the experimental data points of the phase shift, see Fig. 5. Both
results of the fits, $\chi^{2}/d.o.f.=2.54$ and $\chi^{2}/d.o.f.=5.41$, are worsened w.r.t. the result in Sec. 3.1 (in which $\chi^{2}/d.o.f.=1.25$). Fig. 5 also shows that the derivative term is necessary to obtain a theoretical curve
which is in qualitative agreement with the experiment. \begin{figure}[h]
\begin{center}
\includegraphics[width=0.85 \textwidth] {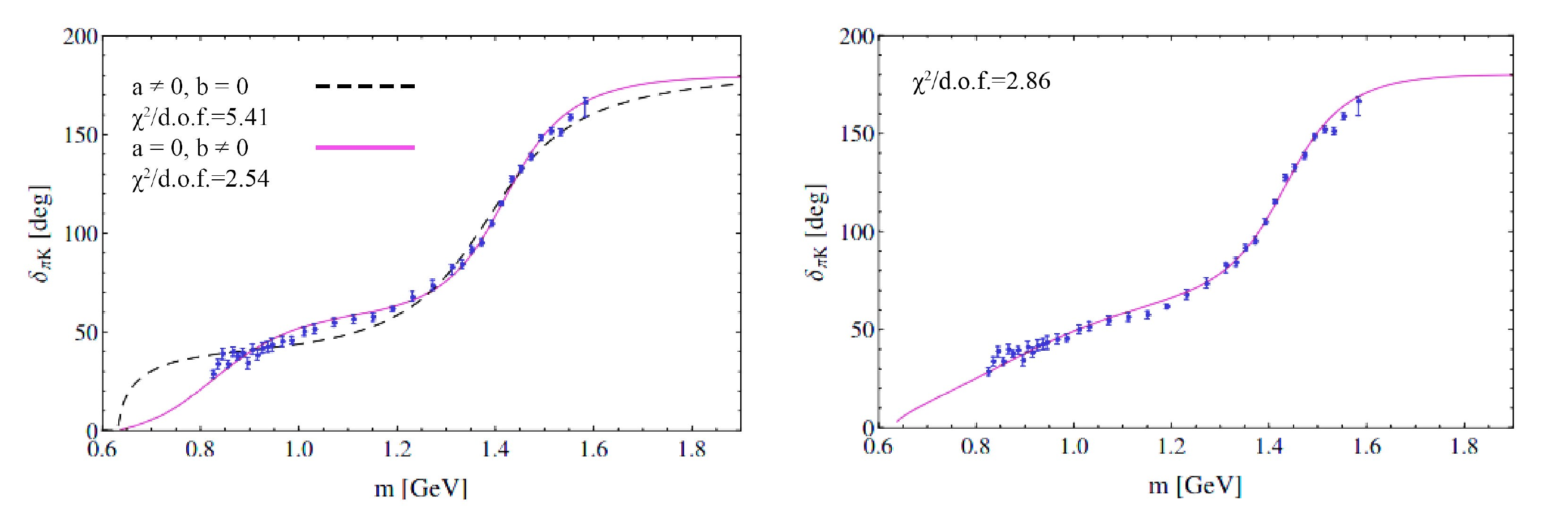}
\end{center}
\caption{Fit results to the experimental points of the phase shift performed
for different types of Lagrangian (left panel) and for modified form factor
(right panel).}%
\end{figure}As a next step, we also test a different form of the form factor.
Namely, although the Gaussian form used in Sec. 3.1 is typically used in such
studies, there is no fundamental reason behind it. Therefore, we have repeated
the fit with:
\begin{equation}
F_{\Lambda}(m)=e^{-2k^{4}(m)/\Lambda^{4}}\text{ .}%
\end{equation}
The result of the fit is presented in Fig. 5. It turns out that a qualitative
agreement is obtained, but the $\chi^{2}$ is clearly worsened: $\chi
^{2}=2.86.$ For the existence and positions of the poles for all these cases,
we refer to \cite{our}. \newpage

\section{Conclusions}

In this work we have studied the properties of the scalar kaonic resonances
$K_{0}^{\ast}(1430)$ and $K_{0}^{\ast}(800)$ according to the formalism
described in Ref. \cite{our}. The original Lagrangian contains a single scalar
field, but because of hadronic quantum loops two poles emerge in the complex
plane once that the parameters of the model are fitted to pion-kaon scattering
data: one pole corresponds to the predominantly quark-antiquark resonance
$K_{0}^{\ast}(1430),$ while the other one is an additional (non
quark-antiquark) companion pole which corresponds to $K_{0}^{\ast}(800).$
While the former survives in the large--$N_{c}$ limit, the latter disappears.
In the spectral function there is only a single peak, roughly corresponding to
$K_{0}^{\ast}(1430);$ in the low-energy domain there is a small enhancement
which is related to $K_{0}^{\ast}(800)$.

In the future, it will be interesting to apply these techniques to other
enigmatic resonances, such as $D_{S0}^{*}(2317)^{\pm}$ and some of the newly
discovered $X,$ $Y,$ and $Z$ states \cite{chen}. \newline\newline\newline

\end{document}